\documentclass[galaxies,article,accept,pdftex,moreauthors]{Definitions/mdpi} 

\firstpage{1} 
\pubvolume{1}
\issuenum{1}
\articlenumber{0}
\pubyear{2024}
\copyrightyear{2024}
\datereceived{ } 
\daterevised{ } % Comment out if no revised date
\dateaccepted{ } 
\datepublished{ } 
\hreflink{https://doi.org/} % If needed use \linebreak
\pdfoutput=1 % Uncommented for upload to arXiv.org
\Title{Semi-empirical Estimates of the Cosmic Planet Formation Rate}

\TitleCitation{Cosmic Planet Formation Rate}

\Author{Andrea Lapi$^{1,2,3,4}$*\orcidA{}, Lumen Boco$^{1,2,3}$\orcidB{}, Francesca Perrotta$^{1}$\orcidC{}, Marcella Massardi$^{4,1}$\orcidD{}}

\AuthorCitation{Lapi, A.; Boco, L.; Perrotta; Massardi}

\address{$^{1}$ \quad Scuola Internazionale Superiore Studi Avanzati (SISSA), Physics Area, Via Bonomea 265, 34136 Trieste, Italy; lapi@sissa.it (A.L.); lboco@sissa.it (L.B.); perrotta@sissa.it (F.P.);  massardi@ira.inaf.it (M.M.)\\
$^{2}$ \quad IFPU-Institute for Fundamental Physics of the Universe, Via Beirut 2, 34014 Trieste, Italy\\ $^{3}$ \quad INFN-Sezione di Trieste, Via Valerio 2, 34127 Trieste,  Italy\\ $^{4}$ \quad IRA-INAF, Via Gobetti 101, 40129 Bologna, Italy\\}

\corres{Correspondence: lapi@sissa.it}

\abstract{We devise and exploit a data-driven, semi-empirical framework of galaxy formation and evolution, coupling it to recipes for planet formation from stellar and planetary science, to compute the cosmic planet formation rate, and the properties of the planets' preferred host stellar and galactic environments. We also discuss how the rates and  formation sites of planets are affected when considering their habitability, and when including possible threatening sources related to star formation and nuclear activity. Overall, we conservatively estimate a cumulative number of some $10^{20}$ Earth-like planets and around $10^{18}$ habitable Earths in our past lightcone. Finally, we find that a few $10^{17}$ are older than our own Earth, an occurrence which places a loose lower limit a few $10^{-18}$ to the odds for a habitable world to ever hosting a civilization in the observable Universe.} 


\begin{document}

\section{Introduction}\label{sec|intro}

One fascinating issue in modern astrophysics concerns the search for the cradles of life: which galaxies and environments along the history of the Cosmos are most likely to host planets, and in particular habitable ones (e.g., with liquid water on their surfaces and a stable atmosphere)? Can these sites resist to environmental threatening from nearby sources (e.g., energy outbursts from supernovae) for sufficiently long times so as to allow the development of life as we know it? 

Although with considerable uncertainties, nowadays we can attempt to answer these questions in a quantitative way. On the one side, the inventory of planets outside the solar system has attained a remarkable number exceeding $5000$ confirmed ones and $7000$ when including all plausible candidates as of August 2024 (see \texttt{https://exoplanet.eu/home/} and \texttt{https://exoplanetarchive.ipac.caltech.edu/}). Moreover, the investigation of the physical properties pertaining to the host stars and environments has allowed to statistically characterize their occurrence as a function, e.g., of the star type/mass and of the heavy element abundance \cite{Lineweaver2001,Fischer2005,Sousa2008,Johnson2012,Gaidos2014,Buchhave2012,Petigura2018,Thompson2018,Zhu2019,Lu2020,Zhu2021,Bello2023,Gore2024}. On the other side, observational and theoretical advancements in galaxy formation and evolution have allowed to determine the amount of star formation in galaxies as a function of the environmental properties (most noticeably, stellar mass and metallicity) and redshift \cite{Kewley2008,Mannucci2010,Andrews2013,Zahid2014,Hunt2016,Maiolino2019,Curti2020,Boco2021,Chruslinska2021,Curti2023,Nakajima2023}. These pieces of information can be combined to provide a global view on the planet formation history along cosmic times, and clarify which galactic environments are more favorable for their occurrence. 
The issue of estimating the number of planets (and especially habitable ones) in the solar neighborhood, in the Milky Way and in the whole Universe has been sporadically considered in the past literature via a variety of ab-initio approaches ranging from (semi)analytic models to numerical simulations \cite{Gonzalez2001,Lineweaver2004,Gowanlock2011,Behroozi2015,Gobat2016,Zackrisson2016,Forgan2017,Stanway2018,Whitmire2020,Balbi2020,Madau2023,Boettner2024}. 

In the present work we take a different, yet complementary, semi-empirical route to the problem. Semi-empirical models \cite{Aversa2015,Moster2018,Behroozi2019,Grylls2019,Hearin2022,Drakos2022,Fu2022,Boco2023,Zhang2023}  have recently emerged as useful frameworks that constitute a kind of `effective' approach to galaxy formation and evolution, in that they do not attempt to model the small-scale physics regulating the baryon cycle from first principles (like it is done in ab-initio semi-analytic models or numerical simulations), but marginalize over it by exploiting empirical relations among spatially-averaged, large-scale properties of galaxies (e.g., star formation rate, stellar mass, metallicity, etc.). The value of these models stands in that they feature a minimal set of assumptions and parameters gauged on observations.
Their results are now routinely used as mock data on top of which to build more complex models or even tune numerical simulations. Moreover, by empirically linking different observables, these frameworks can test for possible inconsistencies among distinct datasets, which often occur given the significant observational systematics in galaxy surveys.
Finally, semi-empirical models are particularly helpful when galaxy formation recipes must be coupled with those from other branches of astrophysics (e.g. stellar evolution and planetary science in the present context), and therefore it is worth to minimize the uncertainties/hypotheses at least on the former side by exploiting basic data-driven inputs. The interested reader can find more details and an extensive discussion on semi-empirical models in the dedicated review by \cite{Lapi2024rev}.

The plan of the paper is straightforward: in Section \ref{sec|modeling} we describe our semi-empirical framework; in Section \ref{sec|results} we present our results; in Section \ref{sec|discussion} we discuss the issue of habitability; finally, in Section \ref{sec|summary} we summarize our findings. Throughout the paper we adopt the standard $\Lambda$CDM cosmology \cite{Planck2020}, and the Chabrier's \cite{Chabrier2003} initial mass function (IMF).

\section{Semi-empirical modeling}\label{sec|modeling}

We build a semi-empirical model by combining data-driven quantities related to galaxy formation and evolution with standard recipes for planet formation.

\begin{center}
\begin{figure}[t]
\includegraphics[width=0.8\textwidth]{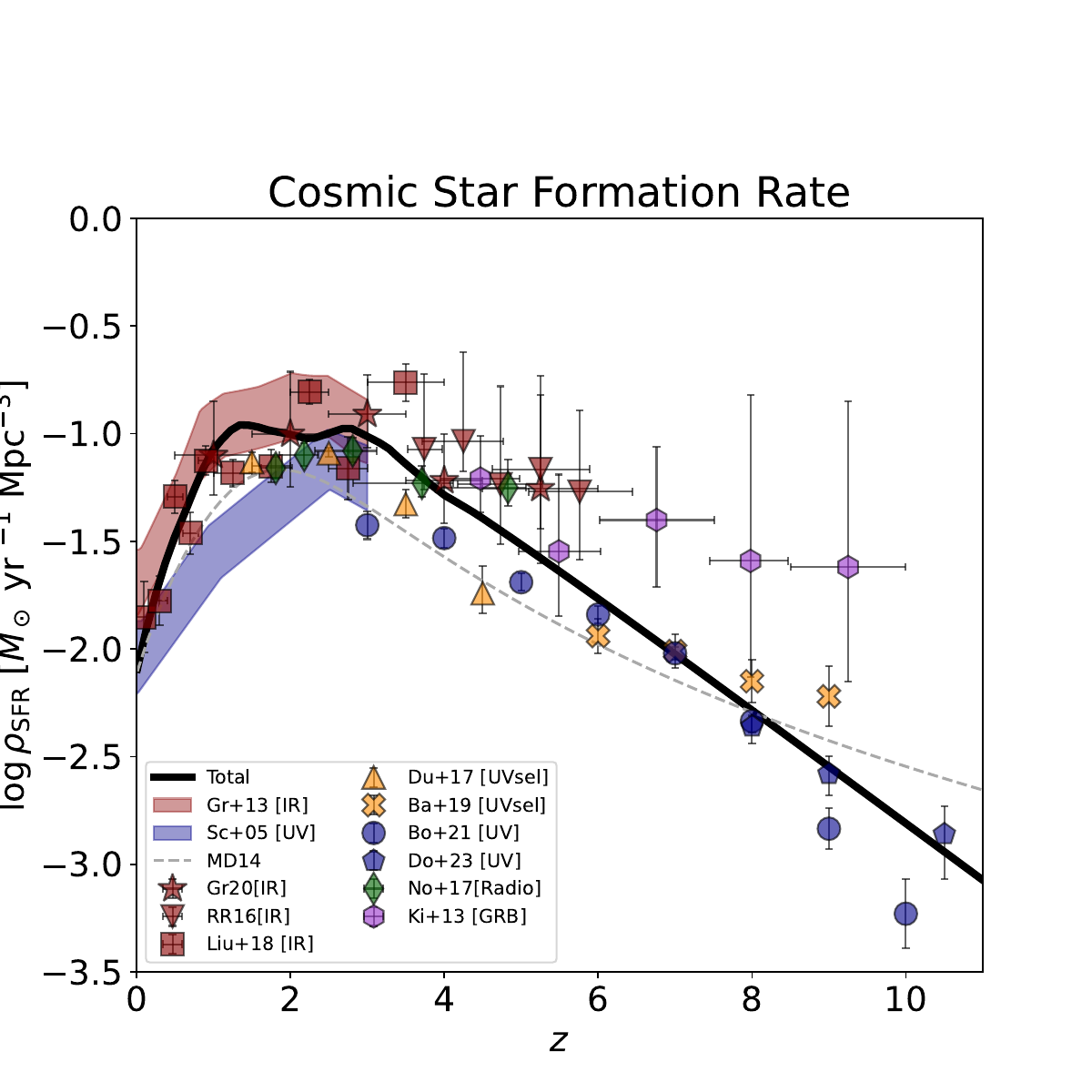}
\caption{The cosmic star formation rate density as a function of redshift. The outcome of our semi-empirical model is illustrated by the black solid line. Data are from \cite{Gruppioni2013} (blue shaded area), \cite{Schiminovich2005} (blue shaded area), \cite{Gruppioni2020} (red stars), \cite{Rowan2016} (red inverse triangles), \cite{Liu2018} (red squares), \cite{Dunlop2017} (yellow triangles), \cite{Bhatawdekar2019} (yellow crosses), \cite{Oesch2018,Bouwens2021} (blue circles), \cite{Donnan2023} (blue pentagons), \cite{Novak2017} (green rhomboids), and \cite{Kistler2013} (magenta hexagons; from GRB). For reference, the dashed grey line shows the classic determination by \cite{Madau2014rev} based on superseded data at $z\lesssim 4$.}\label{fig|SFRD}
\end{figure}
\end{center}

\subsection{Galaxy formation side}

On the galaxy formation side, we start by computing the amount of cosmic star formation rate (SFR) density per galaxy metallicity bin by writing
\begin{equation}\label{eq|galterm}
\frac{{\rm d}\rho_{\rm SFR}}{{\rm d}\log Z} = \int{\rm d}\log M_\star\, \frac{{\rm d}N}{{\rm d}\log M_\star\, {\rm d}V}\, \int{\rm d}\log\psi\, \psi\,\frac{{\rm d}p}{{\rm d}\log \psi}(\psi|M_\star,z)\, \frac{{\rm d}p}{{\rm d}\log Z}(Z|M_\star,\psi,z)~.
\end{equation}
The quantity ${\rm d}N/{\rm d}\log M_\star\, {\rm d}V$ is the latest observational determination of the redshift-dependent galaxy stellar mass function for star-forming galaxies by \cite{Weaver2023}. In addition, ${\rm d}p/{\rm d}\log \psi$ is the distribution of SFR $\psi$ at a given stellar mass $M_\star$ and redshift $z$. Finally, the last quantity ${\rm d}p/{\rm d}\log Z$ is the distribution of metallicity $Z$ at a given stellar mass, SFR and redshift. Note that the inner integral in Equation (\ref{eq|galterm}) represents the average SFR per metallicity bin of an individual galaxy with stellar mass $M_\star$ at redshift $z$; in Section \ref{sec|PFR} we will refer to it as ${\rm d}\langle\psi\rangle/{\rm d}\log Z$.

To compute ${\rm d}p/{\rm d}\log \psi$, we exploit the observational finding that at any given cosmic time the distribution of galaxies in the $\psi$ vs. $M_\star$ plane can be represented as a double log-normal distribution around the galaxy main sequence (MS) and the starburst (SB) population. Therefore one can take
\begin{equation}\label{eq|probsfr}
\frac{{\rm d}p}{{\rm d}\log \psi}(\psi|M_\star,z) = \frac{\mathcal{N}_{\rm MS}}{\sqrt{2\,\pi\,\sigma_{\rm MS}^2}}\, e^{-[\log\psi-\log\psi_{\rm MS}]^2/2\sigma_{MS}^2} + \frac{\mathcal{N}_{\rm SB}}{\sqrt{2\,\pi\,\sigma_{\rm SB}^2}}\, e^{-[\log\psi-\log\psi_{\rm SB}]^2/2\sigma_{SB}^2}~,
\end{equation}
with $\mathcal{N}_{\rm MS}=1-\mathcal{N}_{\rm SB}$ being the fraction of galaxies in the MS, and $\mathcal{N}_{\rm SB}(M_\star,z)$ being the fraction of starburst galaxies by \cite{Chruslinska2021}. Moreover, $\log \psi(M_\star,z)$ is the latest determination of the MS by \cite{Popesso2023}, and $\log\psi_{\rm SB}=\log_{\rm MS}+0.6$ is the typical location of the SB sequence, while $\sigma_{\rm MS}\approx 0.2$ dex is the dispersion of the MS and $\sigma_{\rm SB}\approx 0.25$ dex that of the SB population.

As to ${\rm d}p/{\rm d}\log Z$, observations indicate that it can be described as a log-normal distribution around the fundamental metallicity relation (FMR), i.e.
\begin{equation}\label{eq|probzeta}
\frac{{\rm d}p}{{\rm d}\log Z}(Z|M_\star,\psi,z) = \frac{1}{\sqrt{2\,\pi\,\sigma_{\rm FMR}^2}}\, e^{-[\log Z-\log Z_{\rm FMR}]^2/2\sigma_{\rm FMR}^2} 
\end{equation}
with $\log Z(M_\star,\psi, z)$ the average FMR by \cite{Curti2023} including the high-$z$ correction by \cite{Nakajima2023}, and $\sigma_{\rm FMR}\approx 0.15$ dex being its dispersion.

Note that the cosmic SFR density (the celebrated `Madau' plot) can be just obtained by integrating Equation (\ref{eq|galterm}) over metallicity, to get
\begin{equation}\label{eq|SFRD}
\rho_{\rm SFR}(z)\equiv \int{\rm d}\log Z\, \frac{{\rm d}\rho_{\rm SFR}}{{\rm d}\log Z} = \int{\rm d}\log M_\star\, \frac{{\rm d}N}{{\rm d}\log M_\star\, {\rm d}V}\, \int{\rm d}\log\psi\, \psi\,\frac{{\rm d}p}{{\rm d}\log \psi}(\psi|M_\star,z)~.
\end{equation}
The outcome is shown in Figure \ref{fig|SFRD}, and well agrees with a wealth of observational estimates from data selected in various electromagnetic bands at different redshifts \cite{Gruppioni2013,Schiminovich2005,Gruppioni2020,Rowan2016,Liu2018,Dunlop2017,Bhatawdekar2019,Oesch2018,Bouwens2021,Donnan2023,Novak2017,Kistler2013}. The double peak around the `cosmic noon' at $z\approx 2$ is not to be taken as a realistic feature, since it can be traced back to the detailed shape of the stellar mass function fits by \cite{Weaver2023} that are implemented in Equation (\ref{eq|galterm}). We also stress that stellar mass function determinations are robust out to $z\sim 6$, so at higher redshifts the outcome should be taken with care, and considered as an educated extrapolation albeit it is found in remarkable accord with the latest determinations by the JWST \cite{Donnan2023}.

\subsection{Stellar and planetary side}

On the stellar and planetary side, the crucial ingredient needed in our semi-empirical framework is the occurrence of planets around different type of stars, as a function of the star mass and possibly of other environmental properties like the abundance of heavy elements (see comprehensive textbooks by \cite{Horneck2007,Longstaff2015,Rothery2018,Kolb2021}). In our analysis we will consider M-dwarf stars with masses $m_\star\approx 0.08-0.6\, M_\odot$ and FGK solar-type stars with masses $m_\star\approx 0.6-1.2\, M_\odot$. These two types are thought to be the sites of preferred planet formation, because they are more numerous (for the Chabrier IMF adopted here, FGK-type stars constitute about $12\%$ and M-dwarfs about $80\%$ of the total number) and because more massive stars (except perhaps A-type stars evolving in cool red giants) are thought to hinder or limit planetary formation by a variety of effects (e.g., photo-evaporation of the protoplanetary nebula, stellar winds, SN explosions, etc.).

There are mainly two classes of planets relevant for our analysis.
The first, and most important for obvious reasons, comprises `Earths' (Es). These are routinely defined by a size range $R\approx 0.8-2\, R_\oplus$ and a mass $M\approx 0.5-10\, M_\oplus$, though strictly speaking those in the sub-interval $R\approx 1.25-2\, R_\oplus$ and $M\approx 2-10\,M_\oplus$  are called `super-Earths' \footnote{We stress that the classification of planets in terms of size (or mass) must be taken as merely indicative, since the various intervals are rather arbitrary, and a debate about their very existence is still open (see \cite{Longstaff2015}, Table 13.1; also \cite{Rothery2018,Kolb2021,Boettner2024}).}. Es are almost certainly rocky/metal worlds, with the possibility of locking a certain amount of water on their surface. The second class involves Jupiters with a size range $R\approx 6-150\, R_\oplus$ and masses $M\approx 30-300\, M_\oplus$. Actually most of the planets in this class are found to be gas giants in close orbits to the parent star, an occurrence which enhances their surface temperatures to high values, and therefore makes them `hot Jupiters' (HJs). The relevance of HJs for planetary formation is that they are not born in close-in orbits but have migrated there due to gravitational interactions with other gas giants present in the system, likely destroying any Es in formation during such a gravitational mayhem \cite{Rasio1996,Lin1997,Lineweaver2001,Zackrisson2016,Marzari2022}. 

Finally, it is worth mentioning that planets with sizes and masses in between Earth and Jupiters, i.e. $R\approx 2-6\, R_\oplus$ and $M\approx 10-30\, M_\oplus$, are icy worlds called `Neptunes'. Their number is somewhat uncertain, with a general paucity of hot Neptunes (their atmosphere could have been eroded and these objects could have become smaller, rocky super-Earths) and a relative predominance far away from the parent star beyond the so called snowline, namely the distance from a star beyond which water remains frozen during planetary formation \cite{Suzuki2016,Bourrier2018}. Given the considerable uncertainties in their number and in their impact on the formation/survivability of Es, and given that their hot versions could be transition objects toward super-Es (already accounted for), we do not consider in detail Neptunes in our analysis below. 

\begin{center}
\begin{figure}[t]
\includegraphics[width=0.9\textwidth]{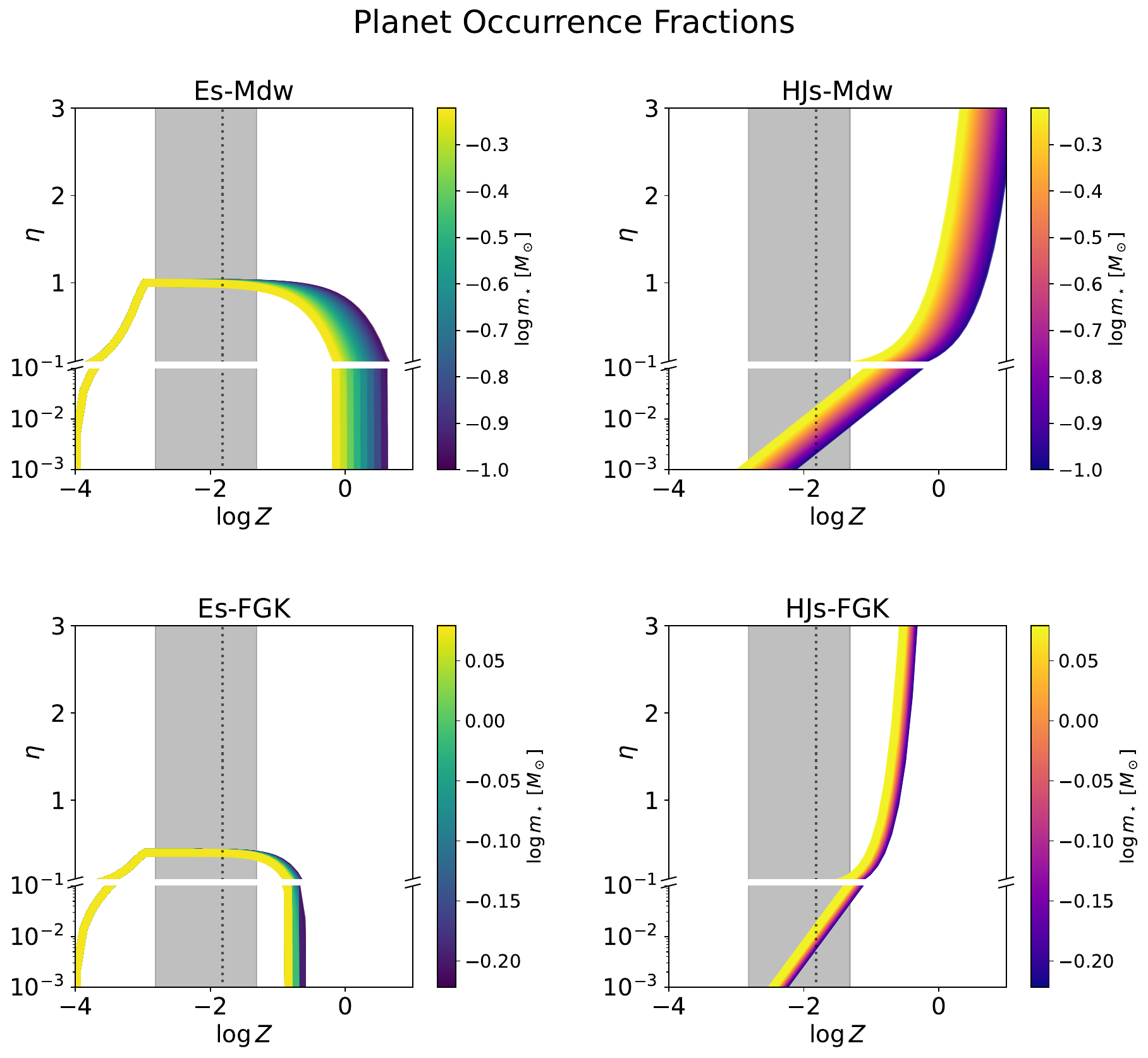}
\caption{The planet occurrence fraction as a function of galaxy metallicity ($x$-axes) and stars' mass (color codes). Top panels refers to M-dwarfs star and bottom panels to FGK stars. Left panels refer to Es and right panels to HJs. In all panels the vertical dotted line highlights the solar metallicity, and the vertical grey shaded area illustrates the metallicity range mostly probed by current observations of exoplanetary surveys. Note that in all panels the scale of the $y-$axis is linear at the top and logarithmic at the bottom, to highlight the behavior of the occurrence for very low metallicities.}\label{fig|POR}
\end{figure}
\end{center}

Observations and theoretical arguments suggest that HJs tend to preferentially form in metal-rich environments (they require a massive $10\, M_\oplus$ iron and silicate kernel to capture their gas envelopes within the lifetime $\lesssim 5$ Myr of the protoplanetary nebula) and around more massive stars \cite{Fischer2005,Petigura2018,Thompson2018,Zhu2019,Zhu2021}. Their occurrence fraction (literally, ratio between the number of planets to the number of stars; note that this quantity in principle is not limited to $1$ from above) is routinely modeled as \cite{Gaidos2014}
\begin{equation}\label{eq|frac_HJs}
\eta_{\rm HJ}(m_\star,Z) = f_{\rm HJ,X}\, 10^{a_{\rm HJ,X}\,{\rm [Fe/H]}}\,(m_\star/M_\odot)^{b_{\rm HJ,X}}
\end{equation}
where $X =$Mdw/FGK is a label that stands for M-dwarf or FGK-type stars, $f_{\rm HJ,FGK}=f_{\rm HJ, Mdw}\approx 0.07$, $b_{\rm HJ,FGK}=b_{\rm HJ, Mdw}\approx 1$, $a_{\rm HJ,FGK}\approx 1.8$ and $a_{\rm HJ,Mdw}\approx 1.06$.
We relate the iron abundance [Fe/H] to the metallicity in solar units  
via the relation [Fe/H] $\approx -0.1+1.182\, \log Z/Z_\odot$ at $z\lesssim 2$ based on Milky-Way element ratios by \cite{Holtzman2015} and via the relation
[Fe/H] $\approx -0.42+\log Z/Z_\odot$ at $z\gtrsim 2$ based on high-$z$ star-forming galaxy estimates by \cite{Stunton2024}. The formation of Es is strongly disfavored by the presence of migrating HJs (see above), while it seems to be somewhat enhanced in M-type dwarfs with respect to FGK-type stars, and to require a minimum threshold in metallicity \cite{Lineweaver2001,Sousa2008,Johnson2012,Buchhave2012,Lu2020,Bello2023,Gore2024}. All these effects are routinely parameterized in terms of the occurrence fraction \cite{Zackrisson2016} 
\begin{equation}\label{eq|frac_Es}
\eta_{\rm E}(m_\star,Z) = \kappa(Z)\,(1-\eta_{\rm HJ}).
\end{equation}
where
\begin{equation}\label{eq|frac0_Es}
\kappa(Z) = f_{\rm E,X}\, \frac{Z-0.0001}{0.001-0.0001}~~~~{\rm for [Fe/H]}\in [-2.2,-1.2],
\end{equation}
while $\kappa=0$ for [Fe/H]$<-2.2$ and $\kappa=1$ for [Fe/H] $>-1.2$; in the above $f_{\rm E,Mdw}\approx 1$ and $f_{\rm E,FGK}\approx 0.4$ apply. 

The planet occurrence fractions described above are illustrated in Figure \ref{fig|POR} as a function of galaxy metallicity and star masses. Note that the occurrence of Earth-like planets is suppressed at very low metallicity (Equation \ref{eq|frac0_Es}) due to the lack of material required to form a rocky body of size $\gtrsim 0.5\, M_\oplus$, and at high metallicity (Equation \ref{eq|frac_Es}) due to the migration effects of HJs whose formation is instead favored. We caveat that these occurrence fractions have actually been checked via observations only in regions relatively close to solar metallicity (vertical dotted lines and grey shaded areas in Figure \ref{fig|POR}), and although extrapolated from state-of-the-art numerical and dynamical simulations they are subject to appreciable uncertainties.

\subsection{Planet formation rate}\label{sec|PFR}

Finally, combining the cosmic SFR density per metallicity bin from Equation (\ref{eq|galterm}) with the metallicity-dependent occurrence fraction of planets from Equations (\ref{eq|frac_HJs}) and (\ref{eq|frac_Es}), we compute the planet formation rate (PFR) as
\begin{equation}\label{eq|PFR}
\begin{split}
{\rm PFR}(z) &= \int{\rm d}\log Z\, \frac{{\rm d}\rho_{\rm SFR}}{{\rm d}\log Z}\,\int_{\rm Mdw/FGK}{\rm d}m_\star\, \phi(m_\star)\, \eta_{\rm Y}(m_\star,Z) = \\
&\\
&= \int{\rm d}\log Z\,\int{\rm d}\log M_\star\, \frac{{\rm d}N}{{\rm d}\log M_\star\, {\rm d}V} \frac{{\rm d}\langle\psi\rangle}{{\rm d}\log Z}\,\int_{\rm Mdw/FGK}{\rm d}m_\star\, \phi(m_\star)\, \eta_{\rm Y}(m_\star,Z)~,
\end{split}
\end{equation}
where $Y=$HJs/Es stands for hot Jupiters or Earths, and the integral over $m_\star$ is made in the range of masses suitable for M-dwarfs or FGK-type stars, where star masses are appropriately weighted by the IMF $\phi(m_\star)$. The first line of the above equation makes clear that the shape of the PFR will differ somewhat from the cosmic SFR. This difference is induced by the metallicity dependence of the planet occurrence fraction, which non-trivially weights the cosmic SFR density per metallicity bin. Note that in the above computation the planet occurrence fraction is assumed to be constant across time; this is reasonable given our present knowledge, but is not granted a-priori and will need to be checked with future data or simulations.
The second line of Equation (\ref{eq|PFR}) recasts the expression in terms of the quantity ${\rm d}\langle\psi\rangle/{\rm d}\log Z$ defined below Equation (\ref{eq|galterm}). This representation makes clear that the PFR per bins of galaxy metallicity $Z$ or galaxy stellar mass $M_\star$ can be derived by avoiding to integrate over the corresponding variable, so highlighting the contribution of galaxies with different properties.

\section{Results}\label{sec|results}

\begin{center}
\begin{figure}[H]
\includegraphics[width=0.8\textwidth]{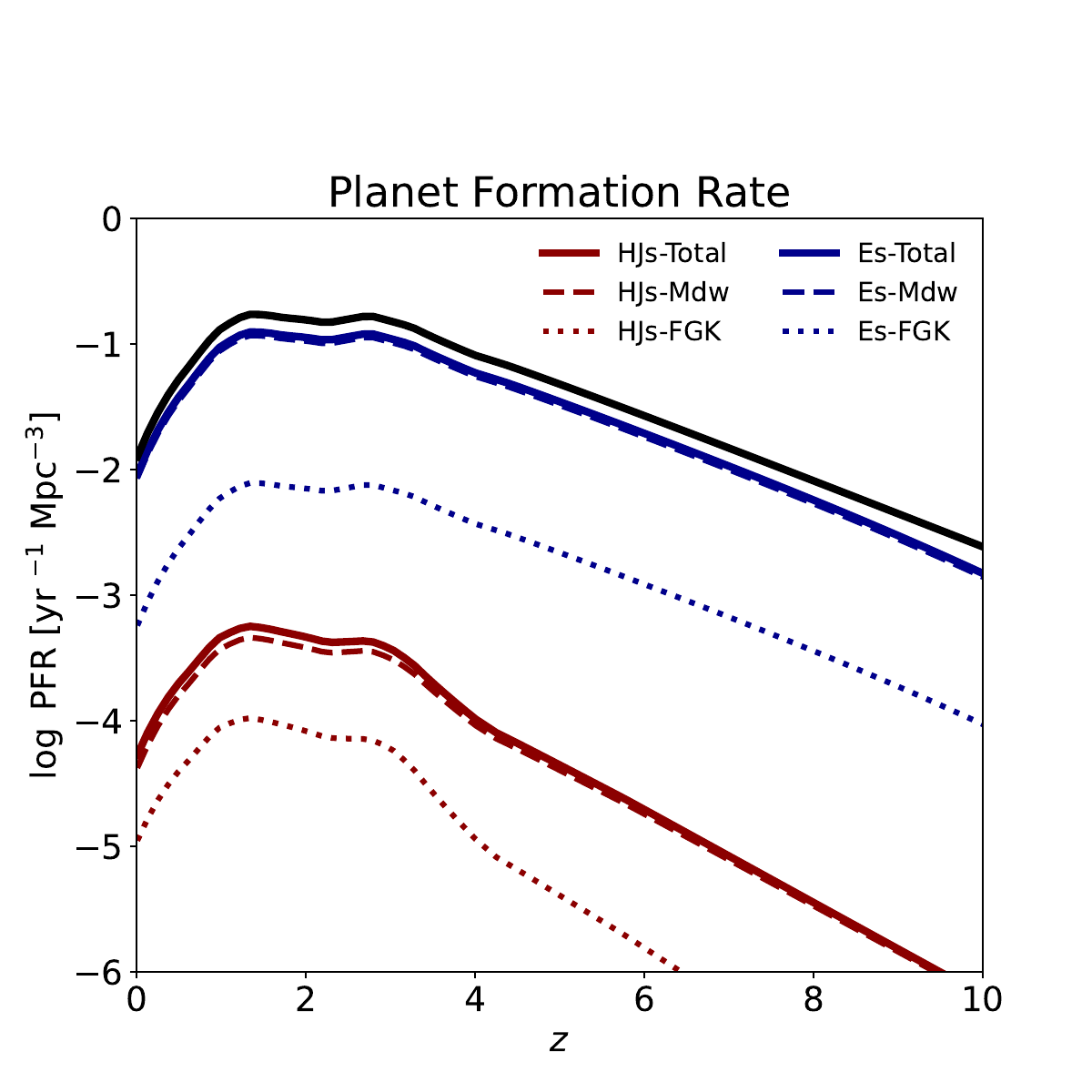}
\caption{The PFR per comoving volume as a function of redshift. Blue lines refer to Es, while red lines refer to HJs. Solid lines illustrates the total rate, while dashed and dotted lines highlight the contribution from planets around M-dwarf stars and FGK-type stars, respectively.}\label{fig|PFR}
\end{figure}
\end{center}

In Figure \ref{fig|PFR} the resulting PFR per comoving volume as a function of redshift $z$ is illustrated. Both for Es and HJs the behavior reflects the cosmic SFR density, with a rise from the present times up to the cosmic noon at $z\approx 1.5$, a rather broad peak there followed by a decline toward higher redshifts. The decline is steeper for HJs, given that their formation requires a rather high metallicity, while it is more gentle for Es whose formation is enhanced by the absence of HJs. The formation rate of Es is about $2$ orders of magnitude higher than for HJs at $z\approx 0$, while at $z\gtrsim 4$ the ratio between the rates increases up to almost $10^3$.
Both for Es and HJs most of the formation occurs around M-dwarfs for two reasons: the number of such low mass stars is appreciably higher in the IMF (approximately $8:1$ number ratio); the planet occurrence fraction (cf. Figure \ref{fig|POR}) around them is higher with respect to FGK-type stars, particularly for Es. 

\begin{center}
\begin{figure}[H]
\includegraphics[width=0.495\textwidth]{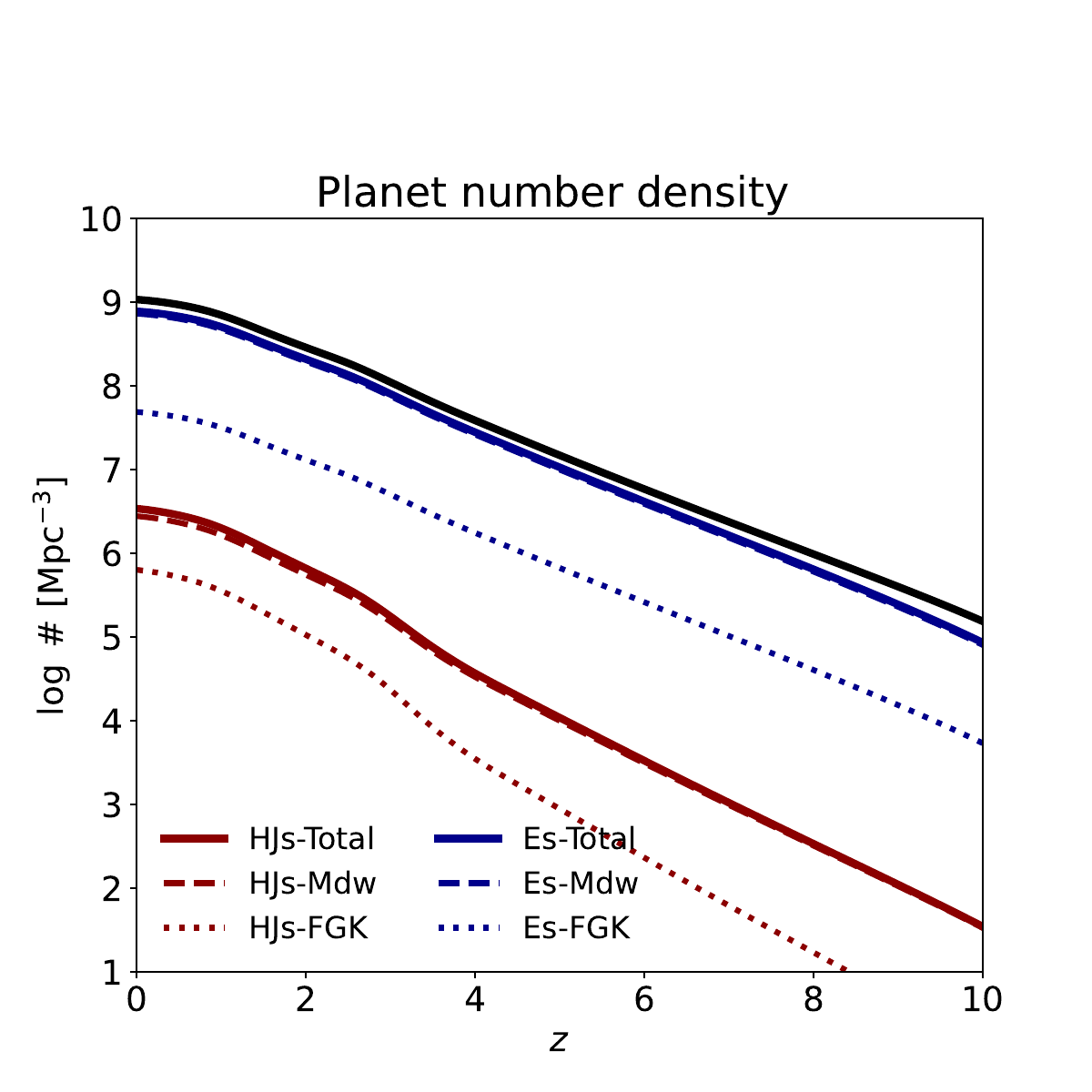}
\includegraphics[width=0.495\textwidth]{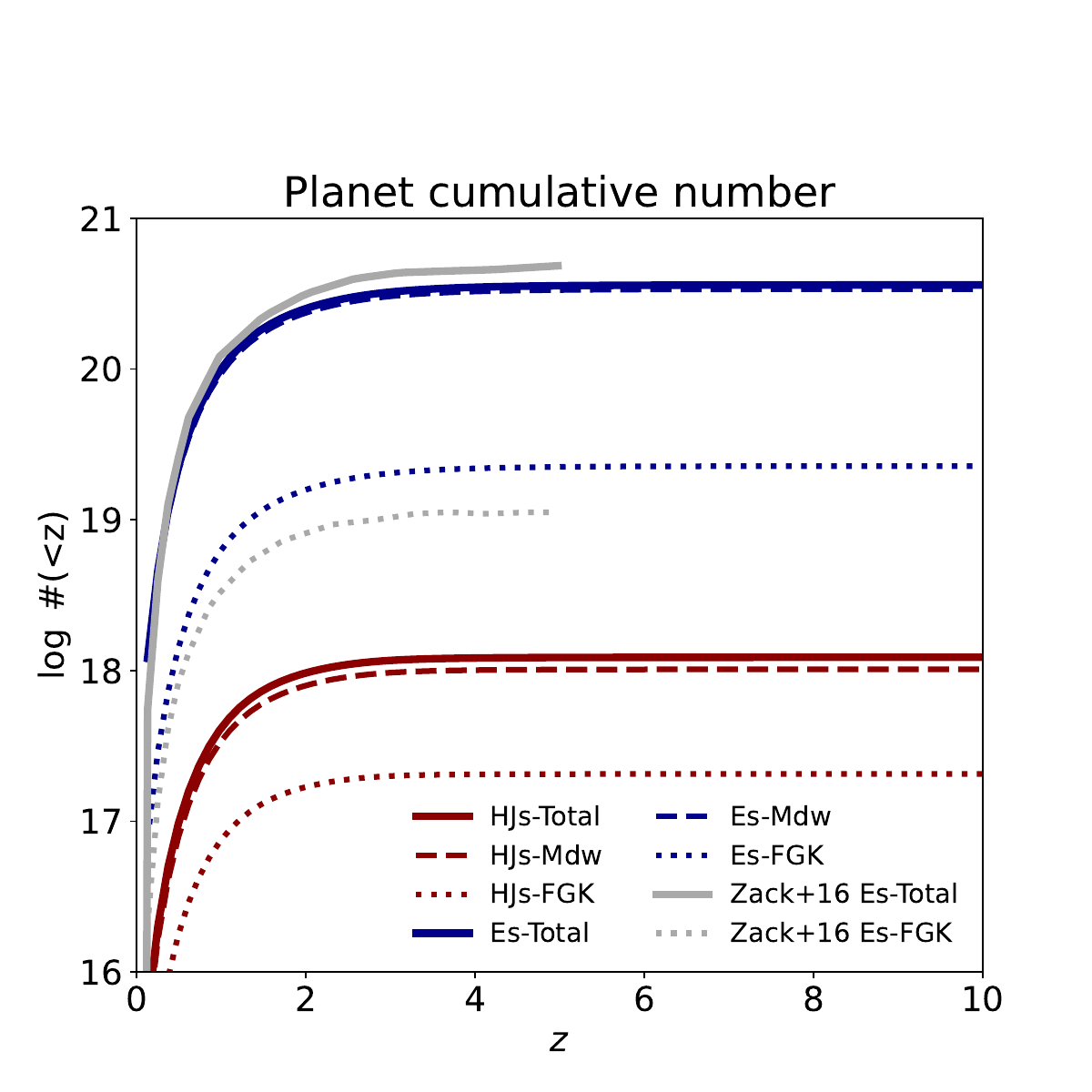}
\caption{The planet number density per comoving volume (left panel) and the cumulative number of planets up to redshift $z$ within our past lightcone (right panel). Linestyles as in Figure \ref{fig|PFR}. In the right panel, the grey lines illustrate the semi-analytic estimates by \cite{Zackrisson2016} for Earth-like planets (solid: FGK+Mdw, dotted: only FGK).}\label{fig|PFRnum}
\end{figure}
\end{center}

Figure \ref{fig|PFRnum} shows the planet number density per comoving volume (left) and the cumulative number of planets up to redshift $z$ (right), obtained by integrating the PFR over cosmic time and volume.
For Es we predict a number density of $10^9$ planets Mpc$^{-3}$ at $z\approx 0$, lowering by a factor $10$ at $z\approx 2$, by a factor $100$ at $z\approx 6$ and by $10^3$ at $z\approx 8$. For HJs we estimate a number density of a few $10^6$ planets Mpc$^{-3}$ at $z\approx 0$, lowering by a factor $10$ at $z\approx 4$, by a factor $100$ at $z\approx 6$ and by $10^4$ at $z\approx 8$.
In terms of cumulative numbers, we expect some $10^{20}$ Es and $10^{18}$ HJs in our past lightcone. This is again mostly contributed by formation around M-dwarfs, while we expect only $10^{19}$ Es and some $10^{17}$ HJs around FGK-type stars. 

It is worth to compare our semi-empirical estimate for the cumulative number of Es to that from the proto-typical semi-analytic model by \cite{Zackrisson2016}, which is illustrated by the grey lines in the right panel of Figure \ref{fig|PFRnum} (solid grey line is the total number of Es, while dotted grey line is the number around FGK stars; cf. right panel of Figure 8 in \cite{Zackrisson2016}). The comparison is instructive because both our framework and the model by \cite{Zackrisson2016} employ very similar prescriptions for planet occurrence fractions, so the difference in the two estimates depends mainly on the galaxy formation side. All in all, the cumulative number of Es is fairly consistent. On a finer detail, with respect to our data-driven framework, the semi-analytic model by \cite{Zackrisson2016} tends to favor the formation of Es (by a factor a few) around M-dwarf stars and disfavor it around FGK-type stars. This is mainly due to the amount of cosmic SFR attributed to galaxies with different metallicities in the two approaches. Note that in semi-analytic models the latter quantity is an output from an ab-initio computation, stemming from the interplay of various complex physical processes (e.g., cooling, star formation, feedback, galaxy merging, etc.). Contrariwise, in our semi-empirical framework it is a data-driven quantity based on the fundamental metallicity relation (see Section \ref{sec|modeling}), and as such the related uncertainty should be minimized or at least reduced.

The comparison of our findings with other literature results is more tricky, since when both the galaxy formation and the planetary science prescriptions are different it is not easy to disentangle the origin of any discordance in the estimates. For example, \cite{Behroozi2015} reports a number of $10^{19}$ Es in the past lightcone (see their footnote 1), which at face value is appreciably different from our result. This is due to a multitude of reasons. On the galaxy formation side, \cite{Behroozi2015} base on a mass-metallicity relation \cite{Maiolino2008} that implies appreciably lower metallicity values toward high redshifts with respect to the fundamental metallicity relation employed by us. This substantially reduces the formation of planets at early cosmic times. On the planetary science side, \cite{Behroozi2015} do not assume that HJs have any adverse effects on the formation of Es, nor treat differently the metallicity dependence and normalization in the occurrence fraction between FGK and M-dwarfs. Moreover, their definition of Es differ somewhat from ours (since they include since the beginning some prior on habitability), in that a more fair comparison would be between our results for Es around FGK stars and their quoted number multiplied by a factor of a few (see \cite{Zackrisson2016}). Doing such a rescaling, their and our results start to be roughly consistent.

\begin{center}
\begin{figure}[H]
\includegraphics[width=0.495\textwidth]{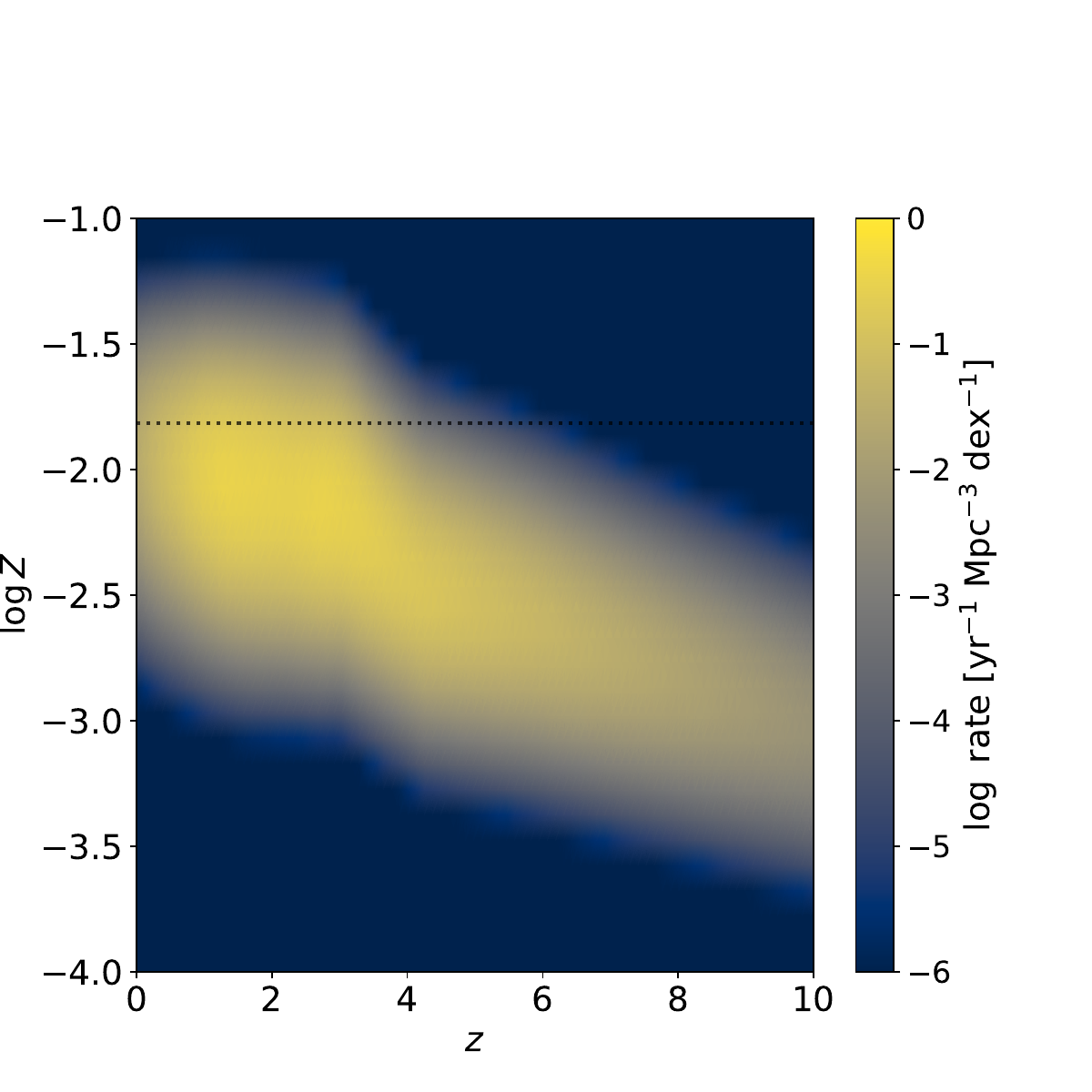}
\includegraphics[width=0.495\textwidth]{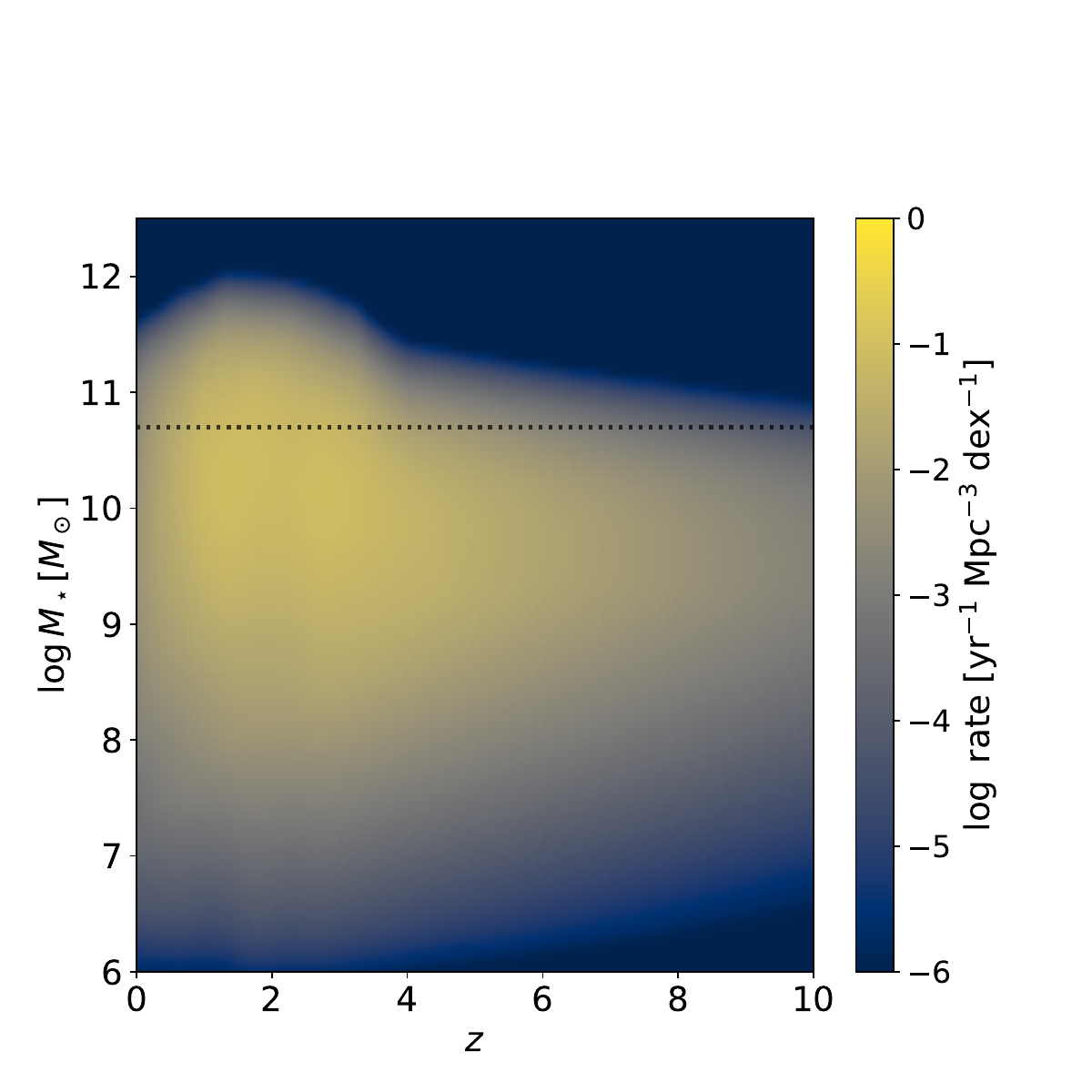}
\caption{The number density of formed stars (color-coded) sliced in galaxy metallicity (left panel) and in galaxy stellar mass (right panel) as a function of redshift. The dotted lines highlight the solar metallicity (left panel) and the stellar mass of a Milky Way-like galaxy (right panel).}
\label{fig|SFR_2D}
\end{figure}
\end{center}

Figure \ref{fig|SFR_2D} displays the number density of formed star as a function of redshift $z$ and metallicity $Z$ (left) or galaxy stellar mass $M_\star$ (right); these are essentially slices of Figure \ref{fig|SFRD} to highlight the properties of the environments where stars are preferentially formed. Most of the stars are formed around solar or slightly subsolar metallicity $Z_\odot$ (dotted horizontal line in the left panel) up to the cosmic noon at $z\sim 2$, and then progressively at smaller metallicity for higher redshifts. However, such a decrease is mild, so that at $z\sim 8$ the typical metallicity where most of the stars are formed amounts to $Z_\odot/10$. Notice that the spread around such values is quite symmetric (in log units), and amounts to $\approx 0.25$ dex. The dependence on galaxy stellar mass is more articulated. Up to $z\sim 2$ most of the stars are formed in galaxies with masses around the Milky-Way value of some $10^{10}\, M_\odot$. This progressively lowers to some $10^{9}$ at higher redshifts. However, the dispersion in these values is very high and asymmetric. At any redshift a substantial amount of stars is formed in galaxies with much lower stellar masses down to $10^8\, M_\odot$.

\begin{center}
\begin{figure}[H]
\includegraphics[width=0.8\textwidth]{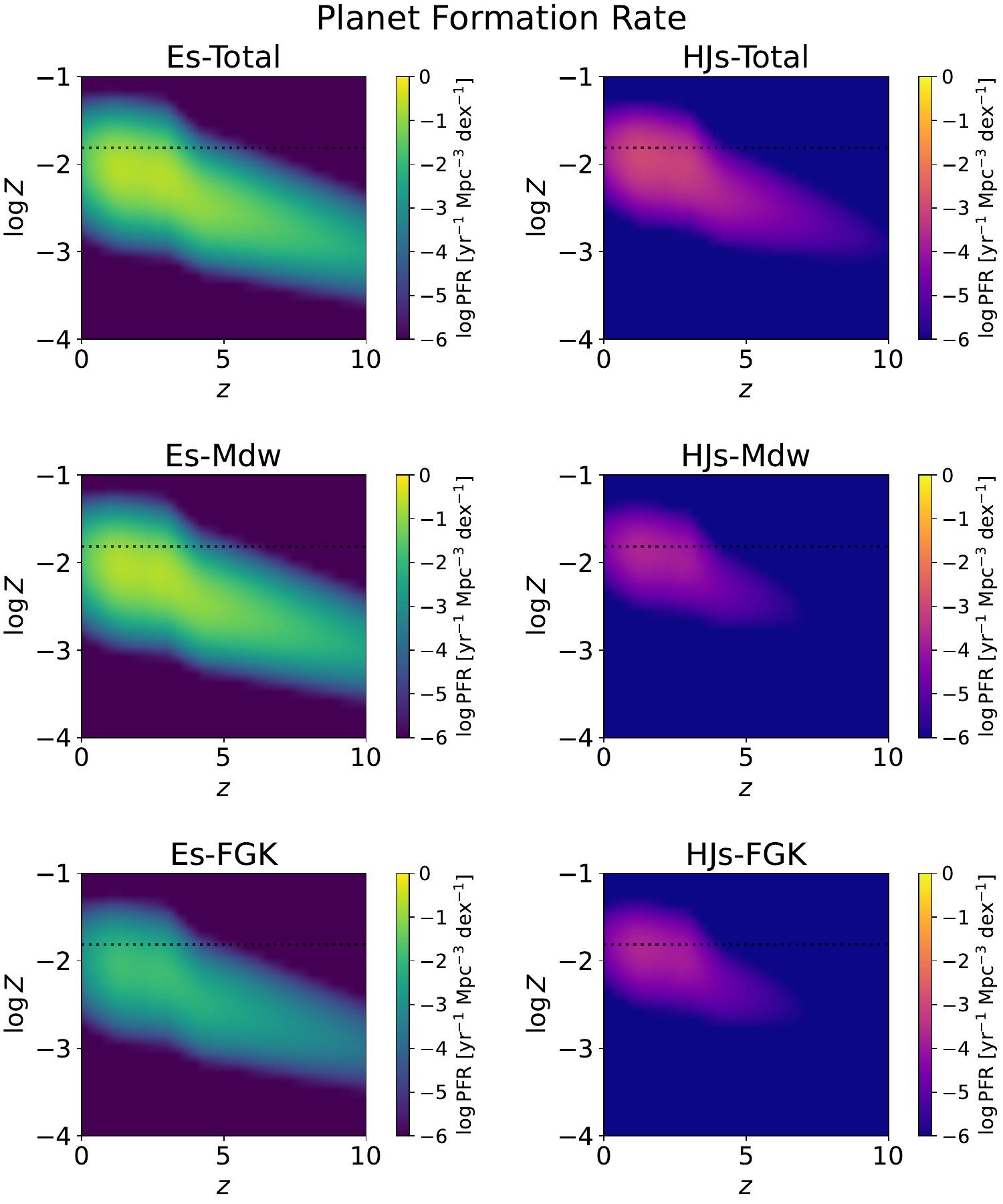}
\caption{The PFR rate (color-coded) sliced in galaxy metallicity as a function of redshift. Left panels are for Es, right panels for HJs; top rows illustrate the total rates, middle rows show the contribution from M-dwarfs stars and bottom rows that from FGK-type stars.}
\label{fig|PFRzeta_2D}
\end{figure}
\end{center}

\begin{center}
\begin{figure}[H]
\includegraphics[width=0.8\textwidth]{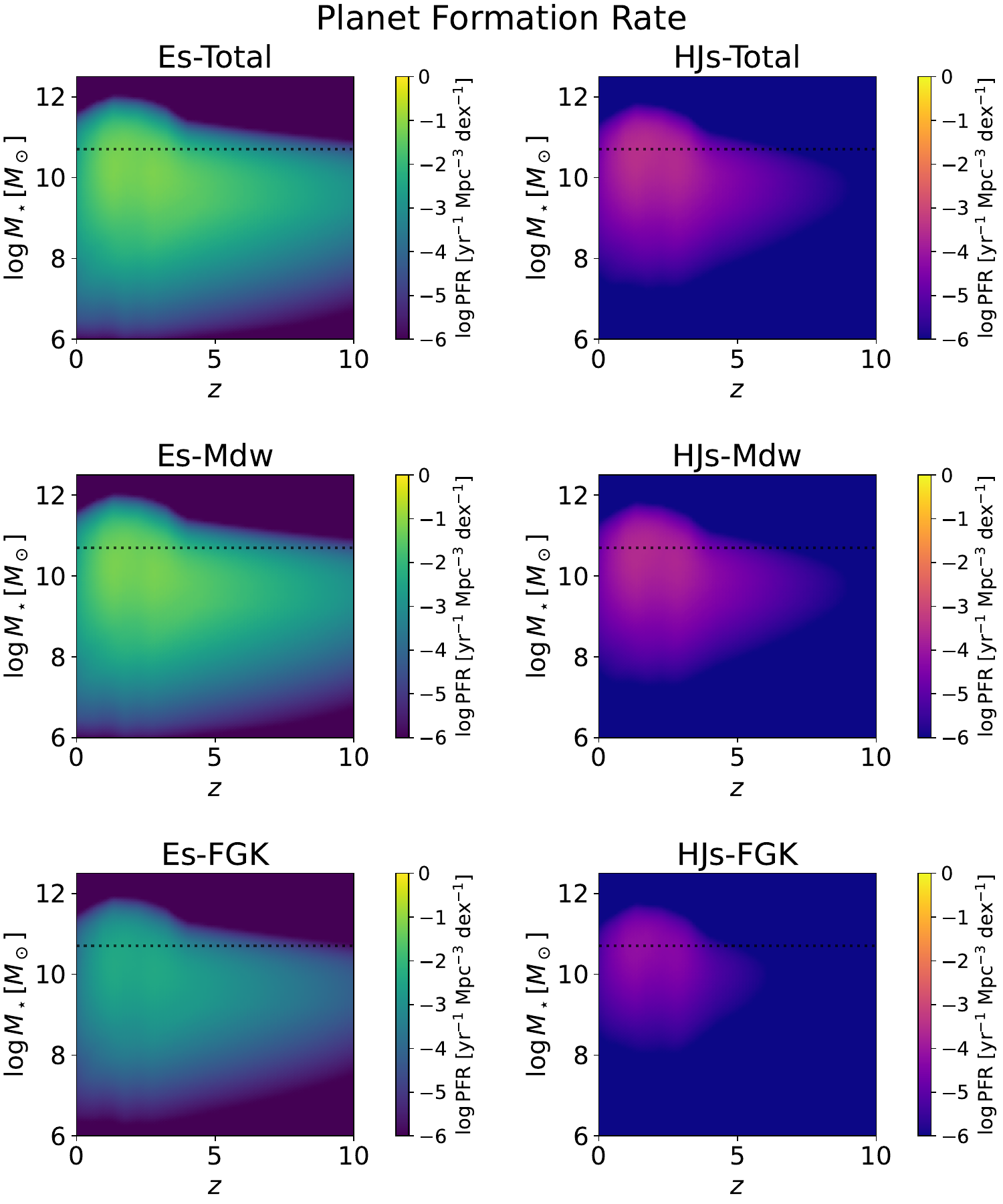}
\caption{The PFR (color-coded) sliced in galaxy stellar mass as a function of redshift. Left panels are for Es, right panels for HJs; top rows illustrate the total rates, middle rows show the contribution from M-dwarfs stars and bottom rows that from FGK-type stars.}\label{fig|PFRmstar_2D}
\end{figure}
\end{center}

The above results are exploited in Figures \ref{fig|PFRzeta_2D} and \ref{fig|PFRmstar_2D} where the PFR is sliced in metallicity and galaxy stellar mass, respectively. Left columns in greenish colors refer to Es, while right columns in reddish colors refer to HJs. Top rows are for the total rates, while middle and bottom rows are for rates around M-dwarfs and FGK-type stars. It is seen that the environments for the formation of Es closely follows those for the host stars both in terms of metallicity and stellar masses (cf. Figure \ref{fig|SFR_2D}). This is not true for HJs, for which both the distribution of  metallicities and of galaxy stellar masses are cut at the low end. This is again related to the requirement of having rather high (solar or supersolar) metallicity for the formation of HJs. On the one hand, this plainly exclude values appreciably lower than solar from the distribution in metallicity. On the other hand, this also cuts galaxies with stellar masses $M_\star\lesssim 10^9\, M_\odot$ since they tend to have low metallicity as prescribed by the observed fundamental metallicity relation (see Section \ref{sec|modeling}).

\section{Discussion: habitability and threatening sources}\label{sec|discussion}

In this Section we discuss how the cosmic PFRs are affected when considering habitability. We caveat the reader that the uncertainties in this context grow much larger, and for this reason it is safe to follow a rather conservative approach based on back-to-the-bones calculations. The occurrence $f_{\rm HEs}$ (i.e., the normalization factor appearing in Equation \ref{eq|frac0_Es}) of habitable Earths (HEs) hosting liquid water on their surfaces and featuring a stable atmosphere/climatic conditions is generally estimated based on some variant of the moist greenhouse criterion \cite{Kasting1993,Kopparapu2014}. 
Currently suggested values are $f_{\rm HEs, FGK}\approx 0.1$ for FGK-type stars and $f_{\rm HEs, Mdw}\approx 0.3$ for M-dwarfs \cite{Petigura2013,Dressing2015,Adibekyan2016,Bryson2021}. 

However, the habitability of planets around M-dwarf stars is highly debated \cite{Luger2015,Sengupta2016,Shields2016,Wandel2018,Chen2019,doAmaral2022,Modi2023}: radiation and flares associated to different evolutionary stages of these stars can erode any stable atmosphere and lead to liquid water evaporation; tidal locking to the host stars as a consequence of synchronous rotation may lead to instability in the climatic conditions. 
All in all, one could expect just a small fraction of HEs around M-dwarf stars. However, given the considerable uncertainties we prefer to follow \cite{Zackrisson2016} and conservatively exclude these planets from the discussion on habitability. On the other hand, using the results of Figure \ref{fig|PFR} it is straightforward for the reader to rescale the formation rate when including M-dwarf stars as possible hosts of habitable planets.

Moreover, the habitability of super-Es planets with masses $1.5-2\, M_\oplus$ is a big question mark \cite{Seager2013,Adibekyan2016,Bergsten2022,Murgas2023}. Development of a gaseous envelope around the big rocky cores of these planets may limit the prospects of hosting life. The empirical finding that the tightness of the orbit depends quite strongly on metallicity makes the number of habitable super-Es highly uncertain. The possibility of having active plate-tectonics, which can stabilize surface temperature, is highly disputed. Therefore we conservatively eliminate super-Es from the balance and focus on proper Es with masses $0.5-1.5\, M_\oplus$ around FGK type stars \cite{Zackrisson2016}. This yields $f_{\rm HE, FGK}\approx 0.04$. Thus at first order one can simply compute the habitable PFR by rescaling the related outcome for Es around FGK stars by a factor $f_{\rm HE, FGK}/f_{\rm E, FGK}\approx 0.04/0.4=0.1$, where $f_{\rm E, FGK}\approx 0.4$ is the occurrence of Es around FGK stars appearing in Equation (\ref{eq|frac0_Es}). In other words, according to these prescriptions, only $10\%$ of the Earth-like planets around FGK stars are considered habitable.

However, this is not the whole story, since events on galactic scales like SNe \cite{Lineweaver2004,Prantzos2008,Carigi2013,Melott2017,Brunton2023,Thomas2023}, (long) GRBs \cite{Thomas2005,Piran2014,Spinelli2023grb} 
supermassive black holes via AGN emission \cite{Gonzalez2005,Balbi2017,Ambrifi2022,Garofalo2023}and tidal disruption events \cite{Pacetti2020} can seriously threaten a planet's habitability. To estimate the effects related to radiation, it is convenient to introduce the critical flux $F_{\rm threat}\approx 100$ kJ/m$^2$ from threatening sources \cite{Gehrels2003,Melott2011,Dartnell2011,Lingam2019,Spinelli2023}, which if emitted in $\gamma$-rays would be able to destroy most of the ozone layer in the atmosphere of HEs, thus exposing the planet surface to harmful UV radiation from the parent star (see references above). Relatedly, one can define a threatening distance from the planet
\begin{equation}
r_{\rm threat} = \left(\frac{E_\gamma}{4\pi\,F_{\rm threat}}\right)^{1/2}
\end{equation}
in terms of the energy output $E_\gamma$ in $\gamma$-rays emitted by the nearby hazardous source. Typical values (see references at the beginning of the paragraph) are $E_\gamma\approx 10^{48}$ erg for SNe II, $\approx 3\times 10^{49}$ erg for SNe I$a$, and $\approx 10^{52}$ erg for GRBs. These yields $r_{\rm threat}\approx 8$ pc for SN II, $\approx 30$ pc for SNI$a$ and about $1$ kpc for GRBs. For AGNs we follow \cite{Gobat2016} and exploit the relation $\log r_{\rm threat}/{\rm kpc}\approx -6+0.7\,\log M_\star/M_\odot$, which yields $r_{\rm threat}\approx$ a few kpc for any reasonable galaxy host with $M_\star\sim 10^{10-12}\, M_\odot$.

The fractional volume threatened by the energy sources reads 
\begin{equation}\label{eq|fthreat}
f_{\rm threat}\approx N_{\rm threat}\times\frac{ 4\, \pi\, r_{\rm threat}^3/3}{M_\star/\rho}\; ;
\end{equation}
here $\rho$ is the average density of the galaxy, that could be roughly taken around $200$ times the matter density of the Universe at the given redshift of the source \cite{Dayal2015}, and $N_{\rm threat}$ is the time-integrated number of threatening events. For SNe and GRBs, the latter quantity depends on the IMF $\phi(m_\star)$, on the SFR $\psi$ and on the stellar age before the explosion $\tau(m_\star,Z)$, in the approximate form
\begin{equation}
N_{\rm threat}\approx N_{\rm norm}\,\psi\,\int_{m_{\star,\rm low}}^{m_{\star,\rm up}}{\rm d}m_\star\; \phi(m_\star)\, \tau(m_\star,Z)\;.
\end{equation}
Specifically, we adopt a Chabrier \cite{Chabrier2003} IMF and compute stellar ages from the evolutionary tracks of the \texttt{PARSEC} code \cite{Bressan2012}. In the above $N_{\rm norm}$ is a normalization factor that takes into account the effects of binary fractions, efficiency in triggering the explosion, delay times, beamed emission, etc. Being the latter very uncertain quantities, we set the normalization factor to the estimates obtained by fitting the cosmic rates of transient events \cite{Gabrielli2024}. For type II SNe $N_{\rm norm}=1$, $m_{\star,\rm low}\approx 8\, M_\odot$ and $m_{\star,\rm up}\approx 60\, M_\odot$ apply. For type I$a$ SN $N_{\rm norm}\approx 0.04$, $m_{\star,\rm low}\approx 2\, M_\odot$ and $m_{\star,\rm up}\approx 8\, M_\odot$ hold. Finally, for long GRBs $N_{\rm norm}\approx 0.1$, $m_{\star,\rm low}\approx 20\, M_\odot$ and $m_{\star,\rm up}\approx 60\, M_\odot$ are adopted. We also implement a threshold in metallicity at $\approx Z_\odot/3$ required for GRB triggering (i.e., $N_{\rm threat}\approx 0$ for larger metallicities), as indicated by observations and simulations \cite{Heger2003}. For AGNs we assume conservatively one outburst ($N_{\rm threat}\approx 1$) per galaxy with stellar mass $M_\star\gtrsim$ some $10^{10}\, M_\odot$.  Finally, the formation rate of HEs that includes threatening effects is computed by inserting the factor $1-f_{\rm threat}$ inside the integral in Equation (\ref{eq|PFR}), taking appropriately into account its possible metallicity dependence. The global effects of independent threatening sources is computed by multiplying the respective $1-f_{\rm threat}$ factors.

\begin{center}
\begin{figure}[t]
\includegraphics[width=0.495\textwidth]{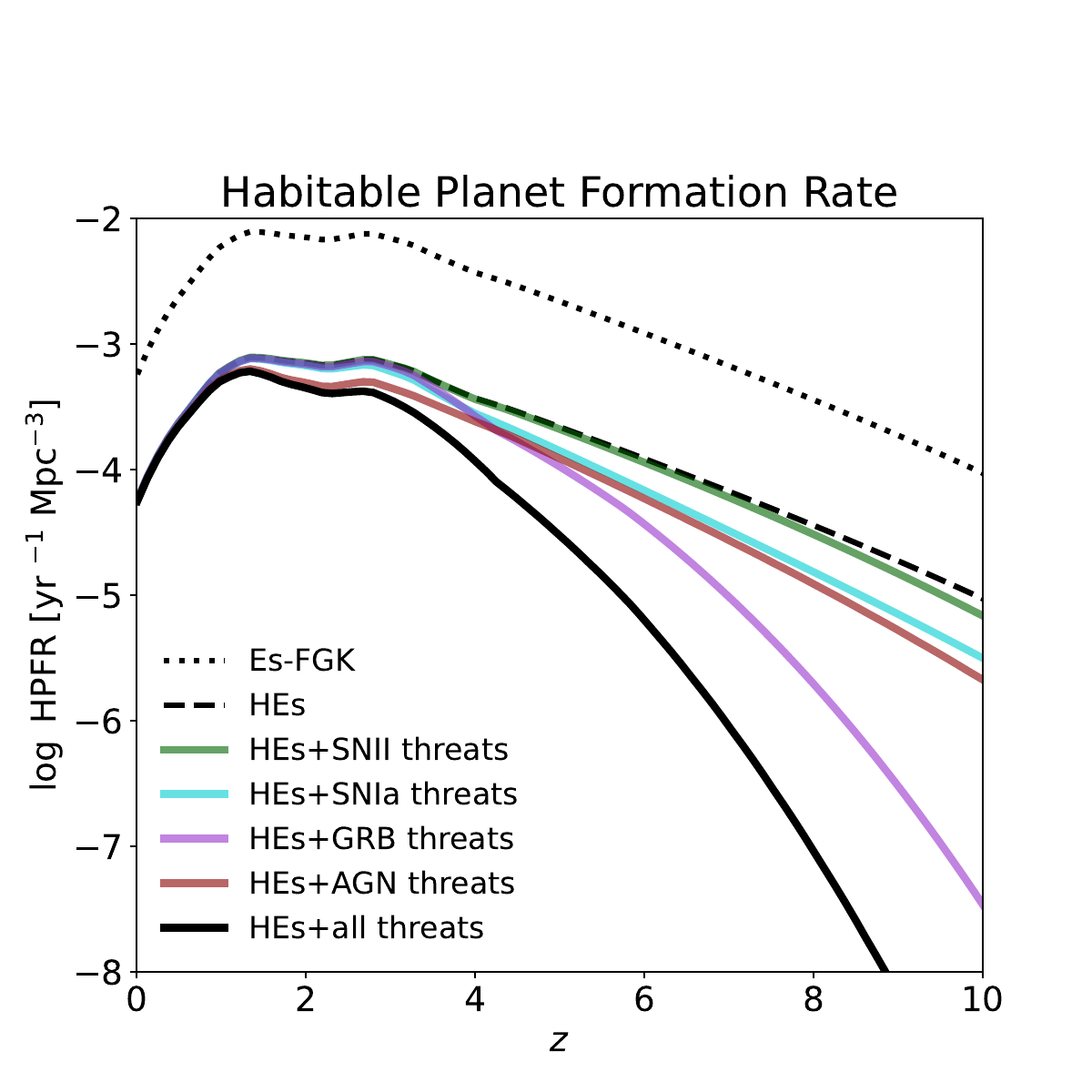}
\includegraphics[width=0.495\textwidth]{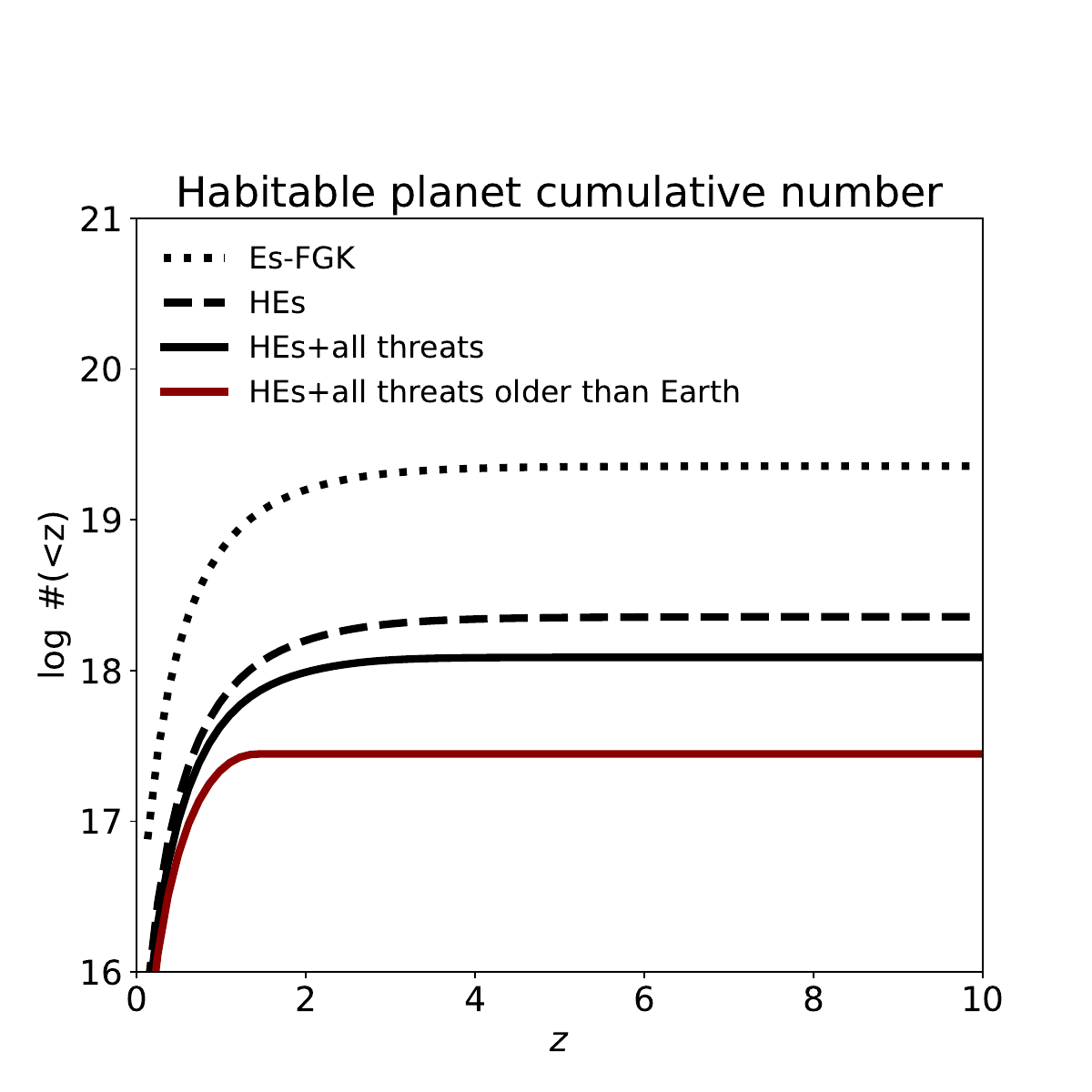}
\caption{Left panel: the habitable PFR as a function of redshift. For reference, black dotted line is the Es formation rate around FGK-type stars. Black dashed line illustrate a conservative estimates to the habitable PFR, taking into account only habitable zone requirements. Colored lines show the reduction in habitable planet rates by various threatening events: SN type 2 (green), SN type I$a$ (cyan), GRB (magenta), and AGN outbursts (red). Finally, black solid line is the global rate when all these threats have been taken into account. Right panel: the cumulative number of habitable planets as a function of redshift in our past lightcone. Linestyles as in left panel. The red line highlight the expected number of planets including all threats and being older than our own Earth (age younger than $4.5$ Gyr).}\label{fig|HPFR}
\end{figure}
\end{center}

\begin{center}
\begin{figure}[t]
\includegraphics[width=0.8\textwidth]{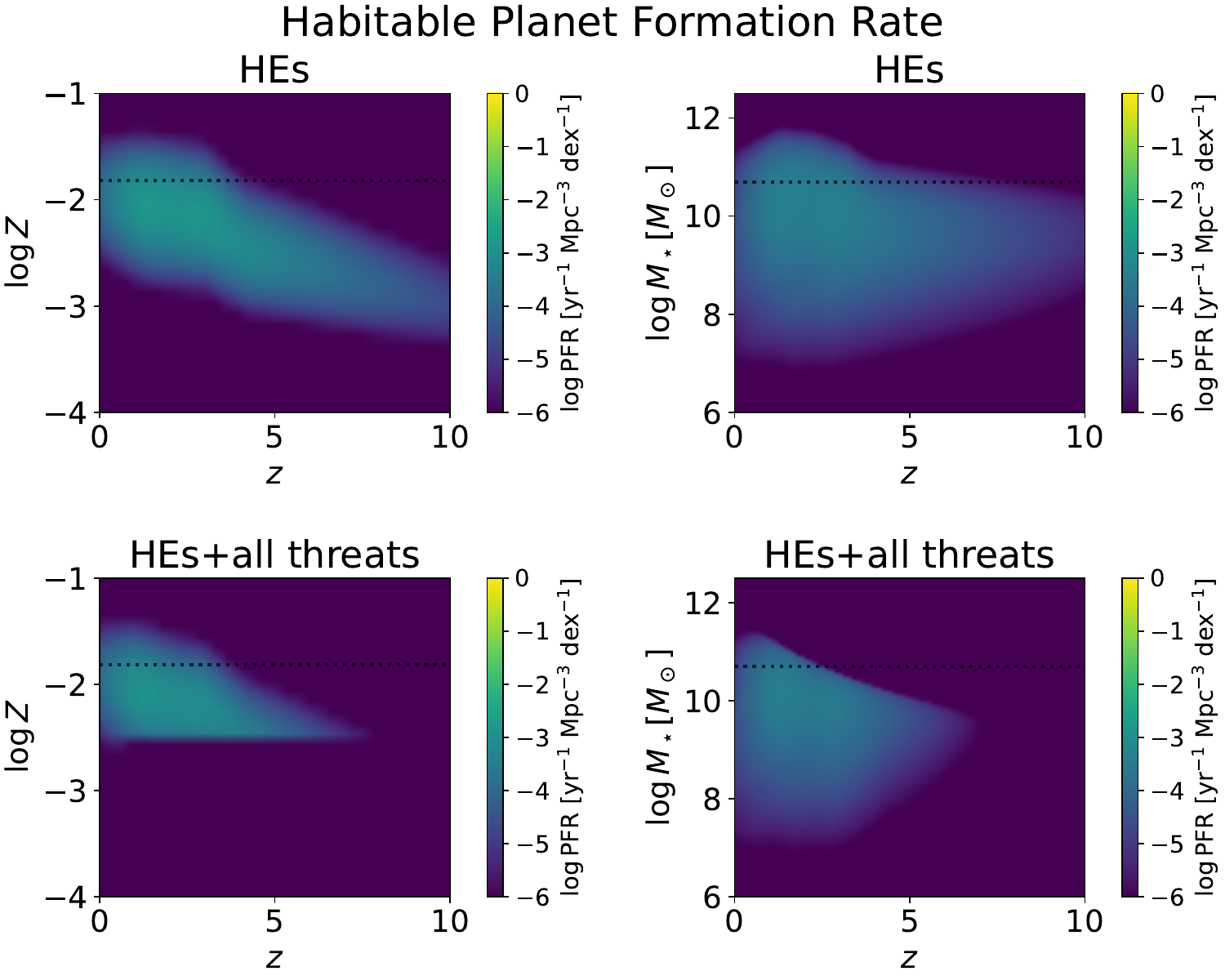}
\caption{The habitable planets formation rate (color-coded) sliced in galaxy metallicity (left panels) and galaxy stellar mass (right panels) as a function of redshift. Top panels refer to habitable zone constraints only, while bottom panels include all threatening events considered in Section \ref{sec|discussion}. Tho dotted lines highlight the solar metallicity (left panels) and the stellar mass of a Milky Way-like galaxy (right panels).}\label{fig|HPFR_2D}
\end{figure}
\end{center}

Results are displayed in Figure \ref{fig|HPFR}, which showcases the habitable PFR and the related cumulative number expected in our past lightcone. For reference, the black dotted lines refer to Es around FGK stars. The black dashed line is the number of HEs with no threatening effects, which would amount to about a few $10^{18}$ in the observable Universe. The black solid line includes all threatening effects considered above, while the colored lines in the left panel refer to individual hazardous sources. The overall effects of threatening sources is to considerably reduce the PFR at high redshift $z\gtrsim 4$, to lower it appreciably around the cosmic noon around $z\approx 2.5$, while mildly affecting it at lower redshifts. The reason behind such a redshift dependence is twofold: (i) at higher $z$ galaxies tend to have higher SFR at given stellar mass, i.e., higher specific SFR $= \psi/M_\star$ entering Equation (\ref{eq|fthreat}) as $f_{\rm threat}\propto \psi/M_\star$; (ii) the density $\rho$ increases at higher $z$, to imply smaller galaxy sizes and hence larger threatened fractional volumes $f_{\rm threat}\propto 1/\rho$ in Equation (\ref{eq|fthreat}). As expected on the basis of their larger energetics, long GRB turn out to be the most hazardous sources, followed by AGNs and type I$a$ SNe, while SN type II have a minor impact. Since the main effects of threatening sources occur at high redshift, when in any case the PFR has substantially decreased with respect to the maximal values at the cosmic noon, the cumulative number of HEs is lowered just by a factor a few, resulting in $\approx 10^{18}$ in our past lightcone. 

Since the estimated age of our own Earth amounts to $\approx 4.5$ Gyr, we can estimate that in our past lightcone the number of planet older than Earth amounts to a few $10^{17}$. The inverse of such a value can be tentatively exploited to claim that the odds for a HEs in the observable Universe to ever hosting a civilization technologically comparable to our own (on assuming similar evolutionary timescales) is around a few $10^{-18}$. Admittedly, this constitutes just a loose lower limit,  which is well within the huge uncertainty range of theoretical estimates based on the Drake equation \cite{Drake1961,Maccone2010,Spiegel2012,Vakoch2015,Loeb2016,Lineweaver2022,Mieli2023}. Note that previous estimates that neglected the metallicity dependence in the planet occurrence fraction, galactic-scale threatening effects, and possible suppression of habitable planet formation around M-dwarf stars ended with much lower odds around $\sim 10^{-24}$ for technological civilizations development \cite{Frank2016,Lingam2021}.

Threatening sources may also alter somewhat the preferred site for the formation of HEs. This is shown in Figure \ref{fig|HPFR_2D}, where the PFR of HEs as a function of host galaxy metallicity (left panels) and stellar mass (right panels)
are compared in case threatening effects are neglected (top panels) or included (bottom panels). The metallicity and stellar mass dependencies of the HEs formation rates are mainly modified at high redshift where threatening effects impact more. In terms of stellar mass, there is a reduction in the formation rates at high masses, due to the presence of more threatening sources associated with star formation and nuclear activity. In terms of metallicity, the most remarkable effect is a considerable reduction of the HEs rates at low metallicity, due to the increased occurrence of long GRBs in metal poor environments.

\section{Summary}\label{sec|summary}

We have devised and exploited a data-driven, semi-empirical framework of galaxy formation and evolution, coupling it to recipes for planet formation from stellar and planetary science, to compute the cosmic planet formation rate, and the properties of the planets' preferred host stellar and galactic environments. Our framework is based on the observed galaxy stellar mass function, star-forming galaxy main sequence and fundamental metallicity relation. These quantities are then combined to the planet's occurrence as a function of metallicity and host star's type and mass, to derive the formation rate of Es and HJs across cosmic times. 

We have determined that both for Es and HJs the behavior of their formation rates features a rise from the present times up to the cosmic noon at $z\approx 1.5$, a rather broad peak there out to $z\approx 3$, followed by a decline toward higher $z$. Such a decline is steeper for HJs, given that their formation requires a rather high metallicity, while it is more gentle for Es whose formation is favored in the absence of HJs. Both for Es and HJs most of the formation occurs around M-dwarf type stars, but an appreciable number is also formed around FGK-type stars.  Overall, we have estimated a cumulative number of some $10^{20}$ Es and around $10^{18}$ HJs in our past lightcone. 

In terms of host galaxy properties, we have found that the majority of the Es are formed around solar metallicity $Z\approx Z_\odot$ up to the cosmic noon around $z\approx 2$, and then progressively at smaller metallicity $\lesssim Z_\odot/10$ toward higher redshifts. The spread around these values amounts to $0.25$ dex. In addition, most of the Es are formed in galaxies with stellar masses around $10^{10}\, M_\odot$ up to the cosmic noon, while toward high redshifts the preferred host galaxies feature stellar masses amounting to some $10^9\, M_\odot$. However, the distribution in stellar masses is largely dispersed and skewed toward low values, in that at any redshift a substantial amount of Es is also formed in galaxies with stellar masses down to $10^8\, M_\odot$. For HJs, the distribution in stellar masses and metallicity are similar but for a cut at the low ends, since these planets require very metal-rich environments to form, which in turn are associated to galaxies with the highest stellar masses.

We have also tentatively discussed how the formation rates and preferred formation sites are modified when considering the planets' habitability, and when including possible threatening sources related to star formation (SNe II/I$a$, long GRBs) and nuclear activity (AGNs). The formation rate of HEs is about one order of magnitude lower than that of Es around FGK stars. This is further reduced, especially at high redshift $z\gtrsim 2$ by threatening sources. Overall, we have conservatively estimated a cumulative number of about $10^{18}$ HEs in the observable Universe. Finally, we have highlighted that this number is reduced to a few $10^{17}$ if considering HEs older than our own Earth, an occurrence which places a loose lower limit of a few $10^{-18}$ to the odds for a habitable world to ever hosting a civilization comparable to ours in the Cosmos.

\vspace{6pt}

\authorcontributions{Conceptualization: A.L.; methodology: A.L., L.B.; validation: L.B., F.P., M.M.; writing: A.L., L.B. All authors have read and agreed to the published version of the manuscript.}

\funding{This work was partially funded from the projects: ``Data Science methods for MultiMessenger Astrophysics \& Multi-Survey Cosmology'' funded by the Italian Ministry of University and Research, Programmazione triennale 2021/2023 (DM n.2503 dd. 9 December 2019), Programma Congiunto Scuole; EU H2020-MSCA-ITN-2019 n. 860744 \textit{BiD4BESt: Big Data applications for black hole Evolution STudies}; Italian Research Center on High Performance Computing Big Data and Quantum Computing (ICSC), project funded by European Union - NextGenerationEU - and National Recovery and Resilience Plan (NRRP) - Mission 4 Component 2 within the activities of Spoke 3 (Astrophysics and Cosmos Observations);  European Union - NextGenerationEU under the PRIN MUR 2022 project n. 20224JR28W "Charting unexplored avenues in Dark Matter"; INAF Large Grant 2022 funding scheme with the project "MeerKAT and LOFAR Team up: a Unique Radio Window on Galaxy/AGN co-Evolution; INAF GO-GTO Normal 2023 funding scheme with the project "Serendipitous H-ATLAS-fields Observations of Radio Extragalactic Sources (SHORES)".}

\dataavailability{N/A}

\acknowledgments{We acknowledge the anonymous referees for useful comments and suggestions. We warmly thank the GOThA team for helpful discussions.}

\conflictsofinterest{The authors declare no conflict of interest.}

\begin{adjustwidth}{-\extralength}{0cm}

\reftitle{References}

\bibliography{bib}

\PublishersNote{}
\end{adjustwidth}
\end{document}